\documentclass{article}

\usepackage[nonatbib, final]{neurips_2024_ml4ps}

\usepackage[utf8]{inputenc} 
\usepackage[T1]{fontenc}    
\usepackage{hyperref}       
\usepackage{url}            
\usepackage{booktabs}       
\usepackage{amsfonts}       
\usepackage{nicefrac}       
\usepackage{microtype}      
\usepackage{xcolor}         

\usepackage{amsmath}
\usepackage{amssymb}
\usepackage{lipsum}
\usepackage{graphicx}
\usepackage{multirow}
\usepackage{caption}
\usepackage{subcaption}
\usepackage{tikz}
\usetikzlibrary{calc}
\usepackage{float}
\usepackage{bbm}
\usepackage{algorithm}
\usepackage{algpseudocode}
\usepackage{cite}

\newcommand{\Cbb}{\mathbb{C}}
\newcommand{\Rbb}{\mathbb{R}}
\newcommand{\Zbb}{\mathbb{Z}}

\def\*#1{\mathbf{#1}}

\newcommand{\Ncal}{\mathcal{N}}
\newcommand{\Pcal}{\mathcal{P}}

\newcommand*{\hermconj}{\text{H}}

\title{Integrating Generative and Physics-Based Models for Ptychographic Imaging with Uncertainty Quantification}

\author{%
  Canberk Ekmekci\thanks{Work done during an internship at Argonne National Laboratory.} \\
  University of Rochester\\
  Rochester, NY 14627 \\
  \texttt{cekmekci@ur.rochester.edu} \\
  \And
  Tekin Bicer \\
  Argonne National Laboratory  \\
  Lemont, IL 60439 \\
  \texttt{tbicer@anl.gov} \\
  \AND
  Zichao Wendy Di \\
  Argonne National Laboratory  \\
  Lemont, IL 60439 \\
  \texttt{wendydi@anl.gov} \\
  \And
  Junjing Deng \\
  Argonne National Laboratory  \\
  Lemont, IL 60439 \\
  \texttt{junjingdeng@anl.gov} \\
  \And
  Mujdat Cetin \\
  University of Rochester\\
  Rochester, NY 14627 \\
  \texttt{mujdat.cetin@rochester.edu} \\
}

\begin{document}

\maketitle

\begin{abstract}
Ptychography is a scanning coherent diffractive imaging technique that enables imaging nanometer-scale features in extended samples. One main challenge is that widely used iterative image reconstruction methods often require significant amount of overlap between adjacent scan locations, leading to large data volumes and prolonged acquisition times. To address this key limitation, this paper proposes a Bayesian inversion method for ptychography that performs effectively even with less overlap between neighboring scan locations. Furthermore, the proposed method can quantify the inherent uncertainty on the ptychographic object, which is created by the ill-posed nature of the ptychographic inverse problem. At a high level, the proposed method first utilizes a deep generative model to learn the prior distribution of the object and then generates samples from the posterior distribution of the object by using a Markov Chain Monte Carlo algorithm. Our results from simulated ptychography experiments show that the proposed framework can consistently outperform a widely used iterative reconstruction algorithm in cases of reduced overlap. Moreover, the proposed framework can provide uncertainty estimates that closely correlate with the true error, which is not available in practice. The project website is available \href{https://cekmekci.github.io/bayesian-inversion-ptychography-generative-priors-website/}{here}.
\end{abstract}

\section{Introduction}
Ptychography is a coherent diffractive imaging technique that enables imaging nanometer-scale features in various applications such as biological imaging~\cite{Klaus2010PtychographyApplicationBiologicalMicroscopy, Wilke2012PtychographyApplicationBacterialCells, Diaz2015PtychographyApplicationCellularUltrastructure, Deng2015PtychographyApplicationGreenAlgae, Lima2013PtychographyApplicationYeast}, battery material characterization~\cite{Sun2021PtychographyBattery,zhao2024suppressing}, integrated circuit imaging~\cite{Holler2017PtychographyApplicationCircuit, Schropp2011PtychographyApplicationCircuit, Deng2017PtychographyApplicationCircuit}, and general materials science~\cite{Esmaeili2013PtychographyApplicationSilk, Trtik2013PtychographyApplicationCement, Chen2013PtychographyApplicationCoating} (see \cite{Rodenburg2019PtychographyBookChapter} for a comprehensive overview of ptychography). At the data acquisition stage, an extended object is scanned by a coherent light beam that follows a trajectory consisting of multiple scan locations, and the resulting diffraction patterns are collected by a far-field detector. At the inversion stage, an iterative reconstruction algorithm such as \cite{Rodenburg2004PIE, Maiden2009ePIE, Thibault2009DM, Maiden2017rPIEmPIE} estimates the underlying ptychographic object from the observed diffraction patterns. Unfortunately, in practice, iterative ptychographic reconstruction methods often require significant amount of overlap between the adjacent scan locations~\cite{Bunk2008OverlapRatePIE}, leading to prolonged acquisition periods and large data volumes.

One may aim to overcome this major challenge by reducing the overlap between neighboring scan locations. However, this results in a more challenging ill-posed inverse problem, leading to a degraded reconstruction performance or to an unstable inversion process when iterative methods involving existing analytical regularizers are used. Motivated by these challenges, several works, e.g., \cite{Aslan2021PtychotomographyPnP, Guan2019PtychographyUNet, Barutcu2022PtychographyGenerativeModelProgressiveTraining, Yang2020PtychotomographyGenerative, Cherukara2020PtychographyPtychoNN, Seifert2024PtychographyGenerative}, have aimed to design deep learning-based ptychographic reconstruction methods that can perform well with sparsely acquired diffraction patterns. Although these methods are capable of achieving state-of-the-art performance in ptychographic inversion, they do not have the ability to provide any information about the inherent uncertainty hidden in the ptychographic inverse problem caused by its ill-posed nature. An attempt to quantify this uncertainty was made by Dasgupta and Di in \cite{Dasgupta2021PtychographyUncertainty}. However, their approach does not enforce any physical data consistency at the inference stage and assumes a total variation~\cite{Rudin1992TV}-based prior on the ptychographic object, which may not capture the complex structure of real-world ptychographic objects.

In this paper, we propose a scalable Bayesian inversion method for ptychography that can perform well with sparsely acquired diffraction patterns while providing uncertainty estimates about the underlying image, capturing the inherent uncertainty in the ill-posed inverse problem. The proposed method achieves high reconstruction quality under sparse data conditions by incorporating prior knowledge about the underlying object into the inversion process with the help of a deep generative model. Next, it leverages the unadjusted Langevin algorithm~\cite{Roberts1996ULA} to quantify the uncertainty on the ptychographic object given the observed diffraction patterns. 
Throughout the iterations of the Langevin algorithm, the proposed method employs a statistical observation model for ptychography, introducing a physical inductive bias and enforcing data consistency at the inference time. We evaluate the proposed method on simulated ptychography experiments. Our results show that the proposed method offers strong reconstruction performance even when the neighboring scan locations have small amount of overlap. Furthermore, we observe that there exists a positive correlation between the uncertainty estimates provided by the proposed method and the true error maps.

\section{Proposed Framework}
\paragraph{Forward and Inverse Problems} Let $\*u \in \Cbb^N$ be the object of interest in vectorized form and $\*f_j \in \Zbb_+^M$ be the diffraction pattern collected at the $j^\text{th}$ scan location by the far-field detector. Under the Poisson measurement noise assumption, the relationship between the object and the measured diffraction patterns is modeled by the following forward problem (observation model):
\begin{equation}
    \*f_j \sim \Pcal \left( \left| \*F \text{diag}(\*w_j)\*S_j \*u \right|^2   \right) \quad \text{for} \quad j = 1, \dots, J,
\label{eq:original-observation-model}
\end{equation}
where $\Pcal(\lambda)$ denotes a Poisson process with the rate $\lambda > 0$; the matrix $\*F \in \Cbb^{M \times M}$ represents the two-dimensional discrete Fourier transform; the operator $\text{diag}: \Cbb^M \to \Cbb^{M \times M}$ maps a given vector to a diagonal matrix whose diagonal entries are given by the input vector; the vector $\*w_j \in \Cbb^M$ is the complex probe illuminating the object at the $j^\text{th}$ scan location; $\*S_j \in \Rbb^{M \times N}$ is a binary matrix extracting the $\sqrt{M} \times \sqrt{M}$ patch located at the $j^\text{th}$ scan location; and $J \in \Zbb_{++}$ is the number of scan locations in the trajectory. For this forward problem, the inverse problem refers to the task of estimating the underlying object from the measured diffraction patterns $\*f \triangleq \begin{bmatrix} \*f_1^\top , \dots, \*f_J^\top \end{bmatrix}^\top$.

\paragraph{Bayesian Inversion with Generative Models for Ptychography} Bayesian inversion aims to solve the inverse problem by calculating the posterior distribution of the underlying object given the observed diffraction patterns, $p_{u | f}(\cdot | \*f)$. This requires defining a likelihood model $p_{f | u}(\*f | \cdot)$ and specifying a prior distribution $p_u$ that represents our \emph{a priori} knowledge about the object. Unfortunately, specifying the prior distribution is challenging in practice since it is hard to express our \emph{a priori} knowledge about the object mathematically while ensuring mathematical tractability. This often results in overly simplified prior distributions, leading to priors that may not capture the complex structure of real-world objects (see \cite{Adler2018DeepBayesianInversion} for an illustrative example).

Inspired by a common approach presented in the literature (e.g., \cite{Bohra2022BayesianInversionGenerativeModels, Patel2022GANMCMCUQ, Dasgupta2024GANFlow, Jalal2021CompressedSensingPosteriorSampling, Whang2021NFForUQ}), we aim to circumvent this problem by representing the prior distribution of the underlying object through a deep latent generative model. In the rest of this paper, we assume that we have access to a trained deep latent generative model $G : \Rbb^Z \to \Cbb^N$, which is trained using a dataset containing many object samples, and that the generative model $G$ is capable of generating the type of objects we would like to reconstruct, i.e., $\*u = G(\*z)$ where $\*z \sim p_z$ is the latent variable of the generator. 

Under this assumption, we substitute the surrogate expression $G(\*z)$ in lieu of the object $\*u$ in the forward problem in \eqref{eq:original-observation-model} and aim to perform Bayesian inversion over the latent variable $\*z$ by generating samples from the posterior distribution of the latent variable given the diffraction patterns, $p_{z | f}(\cdot | \*f)$. By leveraging the Bayes' theorem, we can decompose the posterior distribution as follows:
\begin{equation}
    p_{z | f}(\*z | \*f) = \frac{p_{f | z}(\*f | \*z) p_z(\*z)}{p_f(\*f)},
\end{equation}
where
\begin{equation}
    p_{f|z}\left(\*f|\*z\right) = \prod_{j=1}^J \prod_{m=1}^M \frac{ \left[ | {\*A}_j G(\*z) |^2 \right]_m^{ \left[ \*f_j  \right]_m } }{   \left[ \*f_j  \right]_m! } e^{ - \left[ | {\*A}_j G(\*z) |^2 \right]_m } \quad \text{and} \quad \*A_j \triangleq \*F \text{diag}(\*w_j)\*S_j.
\end{equation}

Unfortunately, the exact calculation of the posterior distribution above is intractable due to the evidence term, ${p_f(\*f)}$. Thus, we leverage a gradient-based Markov Chain Monte Carlo algorithm called unadjusted Langevin algorithm~\cite{Roberts1996ULA} to generate samples from the posterior distribution. The update equation of the corresponding iterative algorithm is given by
\begin{equation}
     \*z^{(k+1)} = \*z^{(k)} + \gamma \nabla_{\*z}  \log p_{f|z}(  \*f | \*z^{(k)} ) + \gamma \nabla_{\*z}\log p_{z}(  \*z^{(k)}) + \sqrt{2 \gamma} \pmb{\varepsilon}^{(k)} 
\label{eq:ula} 
\end{equation}
where $\gamma > 0$ is the step size; $\pmb{\varepsilon}^{(k)} \sim \Ncal(\*0, \*I)$ is a random additive perturbation; and the gradient of the log-likelihood is given by
\begin{equation}
     \nabla_{\*z}  \log p_{f|z}(  \*f | \*z ) = 2 \Re \biggl\{  \*J_G^\hermconj(\*z) \sum_{j=1}^J \*A^\hermconj_j \left[ ({\*A}_j G(\*z)) \odot \left( {\*f_j} \oslash { | {\*A}_j G(\*z) |^2 }  -  \*1 \right) \right] \biggl\},
\label{eq:gradient-log-likelihood} 
\end{equation}
where $\Re$ calculates the element-wise real part of a given complex vector; $\*J_G(\*z)$ is the Jacobian matrix of the generator $G$ evaluated at $\*z$; $(\cdot)^\hermconj$ denotes the Hermitian conjugate; and the symbols $\odot$ and $\oslash$ denote the element-wise multiplication and division operations, respectively. It is worth noting that each iteration of \eqref{eq:ula} utilizes the generative model to represent the prior on the object while utilizing the likelihood function $ p_{f|z}(  \*f | \cdot )$ to enforce data consistency and introduce a physical inductive bias.

After executing this iterative algorithm for $K$ iterations, while calculating the intermediate vector-Jacobian multiplications using automatic differentiation~\cite{Baydin2018AutomaticDifferentiation}, the resulting variables $\{ \*z^{(1)}, \dots, \*z^{(K)} \}$ are used as inputs to the generator to obtain our samples from the posterior distribution of the object. Then, these samples can be used to obtain a reconstructed image, which can be the arithmetic mean of the samples, and an uncertainty map, which can be the pixel-wise standard deviation of the samples.

\section{Experiments and Results}

\paragraph{Experimental Setup} We utilized the MNIST dataset~\cite{Lecun2010MNIST} to create synthetic ptychographic objects. Initially, we resized each image in the dataset to a $64 \times 64$ resolution. Subsequently, we added a constant offset of $0.2$ to each image and normalized the images so that the pixel intensities fell within the range $(0,1]$. Finally, each ptychographic object is constructed by using one of these images for the magnitude and another for the phase. Since the magnitude and phase images of the ptychographic objects were constructed using identical procedures, we trained a single Wasserstein GAN model~\cite{Goodfellow2014GAN, Arjovsky2017WGAN, Gulrajani2017WGANGP} (see \cite{Bohra2022BayesianInversionGenerativeModels} for its desirable theoretical properties relevant to the posterior sampling problem) on both types of images and used the resulting generator twice to obtain the complex-valued generator $G$. 

To simulate the forward problem in \eqref{eq:original-observation-model}, we utilized the Tike software package~\cite{Gursoy2022Tike,yu2021topology,yu2022scalable}. We used a disk probe for the simulations, where the probe size was set to $16 \times 16$, and the radius was fixed to $8$ pixels. The scan locations were created by following a raster scan pattern, where random perturbations to the scan locations were added. To simulate different data acquisition conditions, we repeated these simulations for different overlap ratios and different probe amplitude values. We ran the proposed method for $1000$ iterations, discarding the first $500$ iterations as the burn-in period. We fixed the step size $\gamma$ to $10^{-5}$ and initialized the latent variable such that $G( \*z^{(0)})$ was close to free space in mean-squared error sense. We compared our method with the state-of-the-art iterative reconstruction method rPIE~\cite{Maiden2017rPIEmPIE}, which is frequently used in synchrotron facilities. 

\begin{figure}
\centering
\begin{minipage}[t]{.58\textwidth}
  \centering
    \includegraphics[width=1.00\columnwidth]{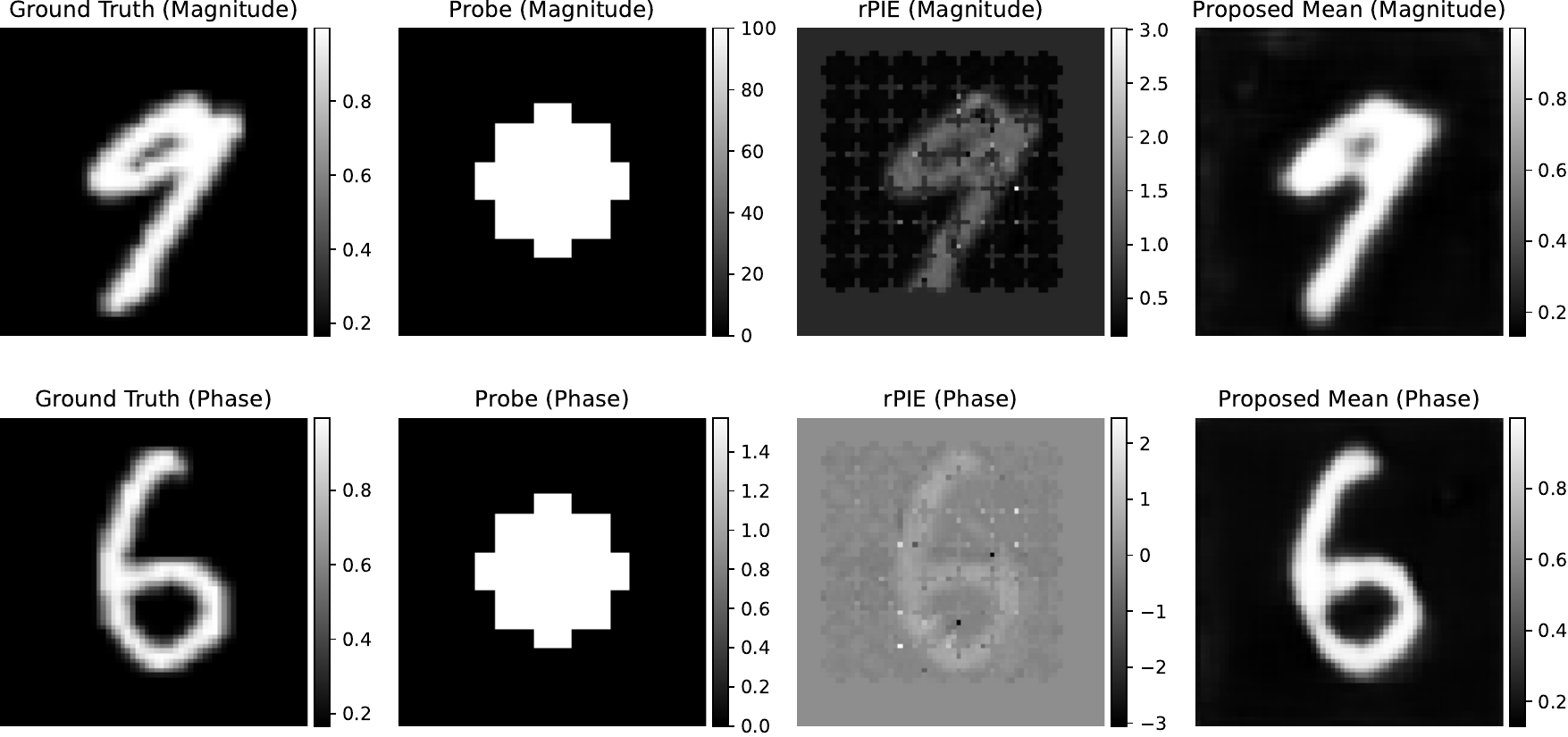}
    \caption{Reconstructions obtained by the proposed method and rPIE~\cite{Maiden2017rPIEmPIE}. The overlap rate is $5\%$, and the probe amplitude is $100$.}
    \label{fig:visual-reconstruction-performance}
\end{minipage}
\begin{minipage}[t]{.37\textwidth}
  \centering
    \includegraphics[width=1.00\columnwidth]{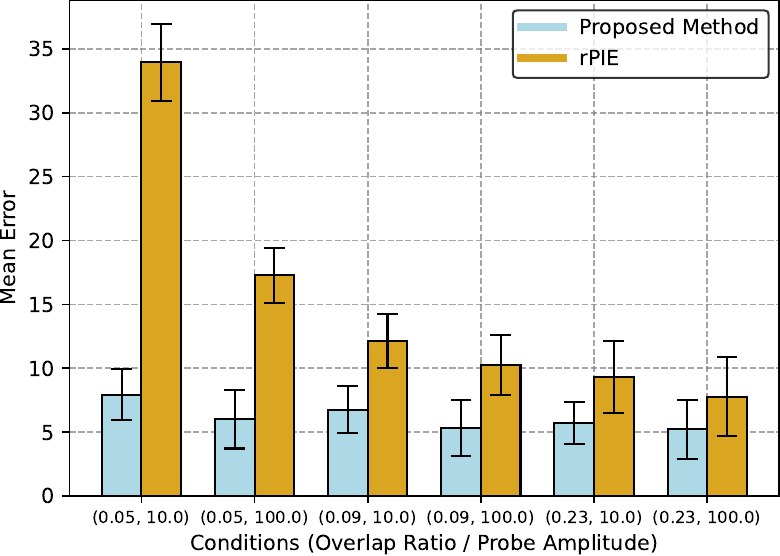}
    \captionsetup{width=.95\linewidth}
    \caption{Reconstruction performance of the proposed method and rPIE~\cite{Maiden2017rPIEmPIE} under various conditions.}
    \label{fig:reconstruction-performance-bar-plot}
\end{minipage}
\end{figure}

\begin{figure}
\centering
\begin{minipage}[t]{.62\textwidth}
  \centering
    \includegraphics[width=1.00\columnwidth]{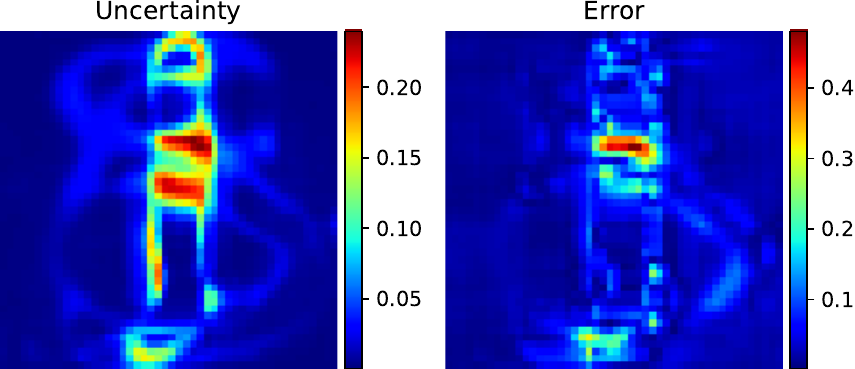}
    \caption{An uncertainty map obtained by the proposed method and the magnitude of the actual error. The overlap rate is $20\%$, and the probe amplitude is $100$.}
    \label{fig:uncertainty-error-correlation-visual}
\end{minipage}
\begin{minipage}[t]{.37\textwidth}
  \centering
    \includegraphics[width=1.0\columnwidth]{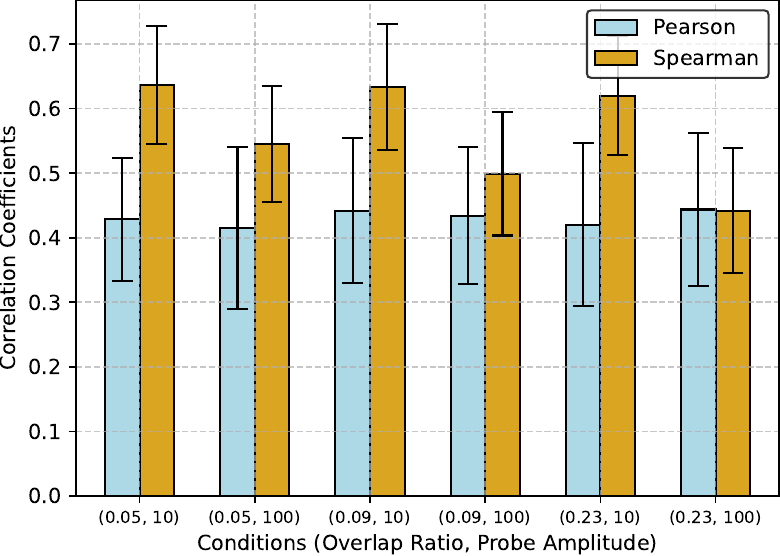}
    \captionsetup{width=.95\linewidth}
    \caption{Correlation between the error and the uncertainty estimates provided by the proposed method.}
    \label{fig:uncertainty-correlation-bar-plot}
\end{minipage}
\end{figure}

\paragraph{Reconstruction Performance} Figure \ref{fig:visual-reconstruction-performance} presents the reconstructed images obtained by the proposed method (see the supplementary material for samples generated from the posterior distribution) and rPIE for a test example, where the overlap ratio is $5\%$, and the probe amplitude is $100$. We observe that both the magnitude and phase images reconstructed by the proposed method exhibit a high degree of visual similarity to those of the ground truth object while the reconstructed images offered by the rPIE algorithm contain severe artifacts. To assess the reconstructions quantitatively, we calculated the $\ell_2-$error~\cite{Albert2020NumericsPhaseRetrieval} between the ground truth object and the mean provided by the proposed method over $100$ test samples. Figure \ref{fig:reconstruction-performance-bar-plot} displays the results. From the figure, we observe that the proposed method consistently outperforms rPIE across all experimental conditions. Moreover, as the overlap ratio and the probe amplitude decrease, the reconstruction performance of rPIE deteriorates significantly, while the proposed framework maintains stable performance, highlighting the advantage of using complex learning-based priors within the reconstruction process.

\paragraph{Uncertainty Estimates} Figure \ref{fig:uncertainty-error-correlation-visual} shows the uncertainty map obtained by the proposed framework and the true error map for a test example, where the overlap ratio is $20\%$, and the probe amplitude is $100$. We observe that the regions with high uncertainty may be used to predict the locations of the potential errors in the reconstructed image without having access to the true ptychographic object. To further evaluate the correlation between the uncertainty maps and the true error maps, we calculated the Pearson and Spearman correlation coefficients for $100$ test samples. Figure \ref{fig:uncertainty-correlation-bar-plot} shows the bar plot of correlation coefficients as a function of the experimental conditions. We observe that Spearman and Pearson correlations are positive across all experimental conditions, indicating a positive relationship between the uncertainty estimates and the actual error. Additionally, we observe that Spearman correlation is consistently higher than the Pearson correlation, suggesting a positive monotonic relationship, rather than a strictly linear one.

\section{Conclusion}
In this paper, we proposed a Bayesian inversion framework for ptychography based on deep generative priors and Markov Chain Monte Carlo sampling. Our simulated ptychography experiments showed that the proposed framework is capable of achieving state-of-the-art reconstruction quality even when the adjacent scan locations have small amount of overlap, where traditional image reconstruction methods often provide erroneous or unstable results. Moreover, we observed that the proposed method is capable of capturing the inherent uncertainty arising in the ptychographic inverse problem, and the provided uncertainty estimates correlate positively with the true error. The main limitation of the proposed method is that it requires training a generative model on the class of ptychographic objects one would like to reconstruct, which may be data intensive for certain experiments conducted at synchrotron radiation facilities. We plan to address this issue in the future by exploring a patch-based adaptation of the proposed method.

\section*{Acknowledgments and Disclosure of Funding}
This work is partially supported by the U.S. Department of Energy (DOE) under Contract No. DE-AC02-06CH11357, including funding from the Office of Advanced Scientific Computing Research (ASCR)'s XSCOPE project and the Laboratory Directed Research and Development Program (Project Number: 2023-0104) at Argonne National Laboratory (ANL). This research used resources of the Argonne Leadership Computing Facility, a U.S. DOE Office of Science user facility at ANL, and is based on research supported by the U.S. DOE Office of Science ASCR Program under Contract No. DE-AC02-06CH11357.

{\small
\bibliographystyle{IEEEtran}
\bibliography{main_bib,bicer}
}

\end{document}